 \documentclass[final,3p,times,twocolumn,sort&compress]{elsarticle}

 \usepackage{graphicx,amssymb,amsmath,amsthm,lineno}
 \usepackage{dcolumn,multirow}
 \usepackage{subfigure}
 \usepackage{threeparttable}

\biboptions{square}

\journal{Brazilian Journal of Physics}

\begin{document}

\begin{frontmatter}



\title{Dynamic Simulation of Random Packing of Polydispersive Fine Particles}


\author[ufsa]{Carlos Handrey Araujo Ferraz\corref{cor1}}
\ead{handrey@ufersa.edu.br}
\author[ufsa]{Samuel Apolin\'ario Marques}
\ead{samuelmarrques@gmail.com}
\cortext[cor1]{Corresponding author}
\address[ufsa]{Exact and Natural Sciences Center, Universidade Federal Rural do Semi-\'Arido-UFERSA, PO Box 0137, CEP 59625-900, Mossor\'o, RN, Brazil}

\begin{abstract}
In this paper, we perform molecular dynamics (MD) simulations to study the two-dimensional packing process of both monosized and random size particles with radii ranging from $1.0 \, \mu m$ to $7.0 \, \mu m$.  The initial positions as well as the radii of five thousand fine particles were defined inside a retangular box by using a random number generator. Both the translational and the rotational movement of each particle were considered in the simulations. In order to deal with interacting fine particles, we take into account both the contact forces and the long-range dispersive forces. We account for normal and static/sliding tangential friction forces between particles and between particle and wall by means of a linear model approach, while the long-range dispersive forces are computed by using a Lennard-Jones like potential. The packing processes were studied assuming different long-range interaction strengths. We carry out statistical calculations of the different quantities studied such as packing density, mean coordination number, kinetic energy and radial distribution function  as the system evolves over time.  We find that the long-range dispersive forces can strongly influence the packing process dynamics as they might form large particle clusters, depending on the intensity of the long-range interaction strength.

\end{abstract}


\end{frontmatter}


\section{Introduction \label{sec:int}}

Random packing of spherical particles have been an important research matter mostly because it can model many complicated processes present in physics and materials engineering. Some applications are found in modeling of ideal liquids~\cite{bernal60,bernal64}, amorphous materials~\cite{finney76,angell81}, granular media~\cite{herrmann95}, emulsions~\cite{pal2008}, glasses~\cite{lois2009}, jamming~\cite{hern2003}, ceramic components~\cite{mcgeary61, hamad90} and densification processes during sintering~\cite{helle85,nair87}. The understanding of the final structure of the particles aggregation is important because its physical properties may depend on the packing features such as packing density, mean coordination number, porosity and radial distribution function (RDF). In particular, the study of the fine particle packing is more difficult to accomplish when compared to the coarse particle packing once long-range interactions between particles, as van der Waals and electrostatic forces, must also be taken into account in the packing formation besides the existing contact forces. In fact, depending on the electrical nature of the involved particles the van der Waals forces are due to instantaneous dipole-dipole interactions (London forces), permanent dipole-dipole interactions (Keesom forces) and/or permanent dipole-induced dipole interactions (Debye forces). Previous works~\cite{yen91, yang2000, cheng2000, jia12} have shown that van der Waals forces can form local particle clusters which hamper the particle rearrangement. Moreover, they can change the packing structural characteristics depending on the particle properties. For example, the packing of micro-sized particles, or smaller, by means of particle deposition is found to be dominated mostly by these long-range dispersive forces~\cite{visser89,israelachvili92}.

Computer simulation based on the so-called distinct element method (DEM)\cite{cundall79} has been used to study fine particle packing. Significant contributions in the study of these processes were given by Yen and Chaki~\cite{yen91}, Cheng {\it et al}~\cite{cheng2000} and Yang {\it et al}~\cite{yang2000} using DEM. Collective rearrangement models~\cite{jodrey85,clarke87,nolan93, he99, kenneth2014} have also been used to simulate particle random packing.  More recently, Jia {\it et al}~\cite{jia12} used DEM to study fine particle packing with Gaussian size distributions. In last decades, most efforts have been employed to study the packing of either monosized particles or particles with a well-known size distribution. However, metal or ceramic powders as well as manufacturing products may be composed by polydispersive particles that not necessarily obey a well-known size distribution. The main advantage of the usage of polydispersive particles in the industry is to allow the increase of density~\cite{konakawa90}
and fluidity~\cite{shapiro92,sudduth93} of the formed compounds. To the best of our knowledge, packing studies concerning random size particles without a well-defined distribution have been rarely reported in the literature. The main aim of this work is to address the problem of the packing of particles with random sizes inside a closed region as well as to study the influence of the long-range dispersive forces during these packing processes.

\begin{figure}[!b]
 \centering
 \includegraphics[scale=0.30]{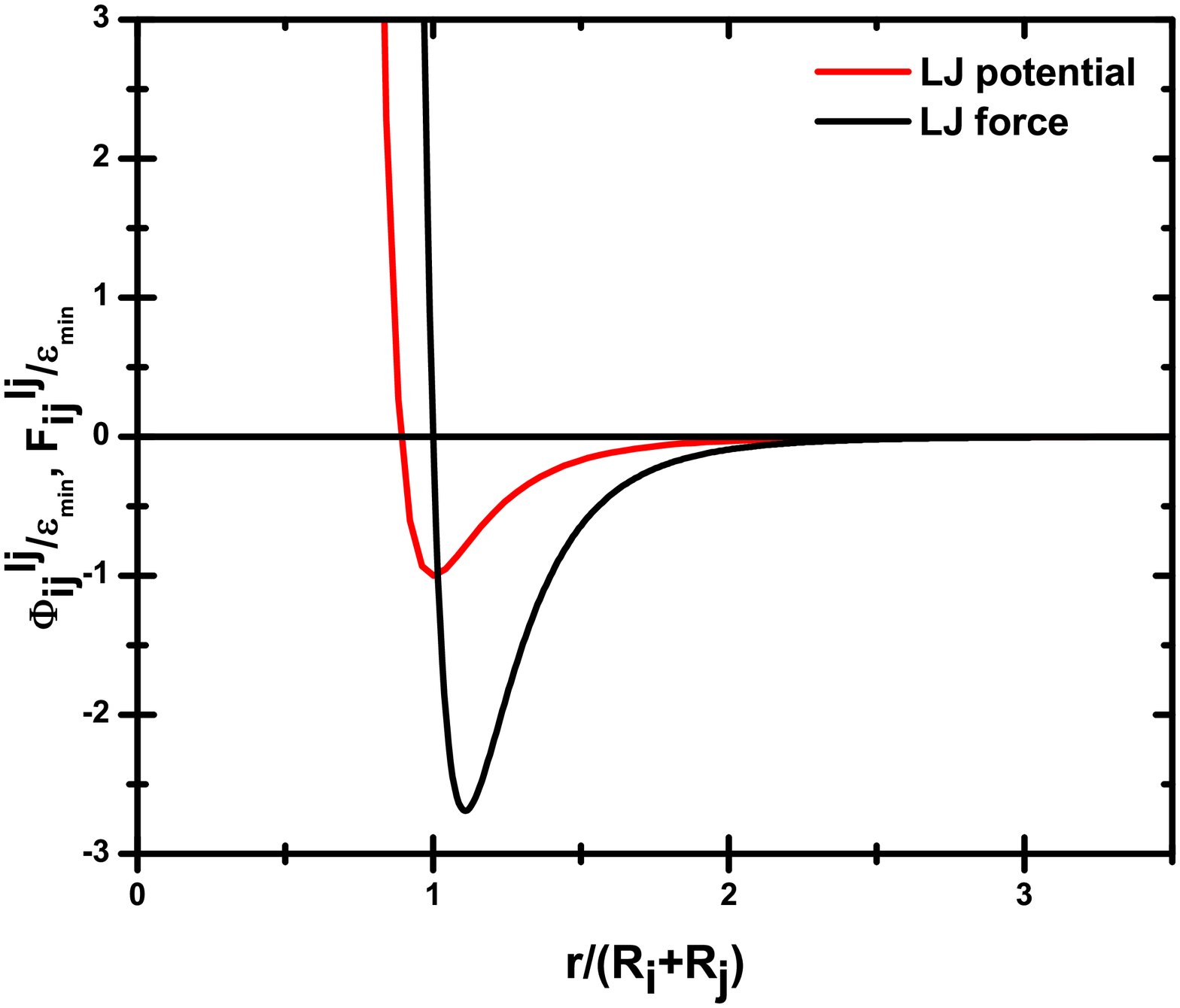}
 \caption{ plot of the LJ potential $\Phi_{ij}^{lj}/\varepsilon_{min}$ (red line) and of the LJ force $F_{ij}^{lj}/\varepsilon_{min}$ (black line) as a function of the relative distance $r/(R_{i}+R_{j})$ between a pair of particles of radii $R_{i}$ and $R_{j}$.} \label{fig:01}
\end{figure}

In this paper, we performed molecular dynamics (MD) simulations to study the two-dimensional(2D) packing process of both monosized and random size particles, that is, polydispersive particles with radii assigned at random. The system was allowed to settle under gravity towards the bottom of a $300 \, \mu m \times 500 \, \mu m$ rectangular box. The initial positions as well as the radii of five thousand non-overlapping particles were defined along the box by using a random number generator~\cite{recipes96}. Both the translational and the rotational movement of each particle were considered in the simulations. In order to deal with interacting fine particles, we take into account both the contact forces and the long-range dispersive forces (van der Waals and electrostatic forces). The contact force results from the deformation of the colliding particles and can be decomposed into two types: normal viscoelastic force and tangential friction force. We account for normal and static/sliding tangential friction forces
between particles and between particle and wall by means of a linear-spring model~\cite{francesco2004}, while the long-range dispersive forces are computed by using a Lennard-Jones like potential~\cite{lennard31}. The validity of this approach is based on the fact that when two microspheres of radius $R$ are separated by a distance $D >> R$, the effective potential ($\Phi$) is analogous to that one between two molecules, i.e., falling off as $\Phi(D)\varpropto -1/D^6$~\cite{israelachvili92, lyklema91}. The packing processes are studied assuming different long-range interaction strengths. We perform statistical calculations of the different quantities studied such as packing density, mean coordination number and time derivative of the kinetic energy as the system evolves over time. For polydispersive particles, a size spectral analysis was employed in order to obtain the RDF of the formed random close-packed structures (RCPS). Apart from more rigorous definitions~\cite{torquato2000}, RCPS is typically defined as a collection of particles packed into its densest possible amorphous configuration.
\begin{table}[!t]
\centering
\small
\begin{threeparttable}
\caption{\label{table:01} Physical parameters used in the simulations.}
\begin{tabular}{lc}
\hline \hline \\
  Parameter\tnote{a} &  Value \\ \hline \\
  Number of particles ($N$) & $5000$ \\
  Particle size ($R$) & $1.0-7.0\;\mu \,m $ \\
  Particle density ($\rho$) & $100/\pi \; Kg/m^{2}$ \\
  Contact normal stiffness ($k_{n}$) &  $1.0 \times 10^{11} \; N/m$ \\
  Contact tangential stiffness ($k_{t}$) & $5.0 \times 10^{8} \; N/m$ \\
  Viscous normal damping ($\gamma_{n}$) &  $15.0  \; Kg/s$ \\
  Viscous tangential damping ($\gamma_{t}$) &  $2.0  \; Kg/s$ \\
  Background friction damping ($\gamma_{b}$) &  $5.0  \; Kg/s$ \\
  Minimum potential energy ($\varepsilon_{min}$) & ($0$; $-5.0$; $-10.0)\;\mu J$ \\
  Static friction coefficient ($\mu_{s}$) & $0.50$ \\
  Dynamic friction coefficient ($\mu_{d}$) & $0.30$ \\
\hline \hline
\end{tabular}
\begin{tablenotes}
      \item[a]{It is assumed that both particles and walls have the same physical parameters.}
    \end{tablenotes}
\end{threeparttable}
\end{table}

The contents of the article are organized as follow. In next section, we describe details of the model and MD simulation background. In section \ref{sec:r}, we  present and discuss the results. Finally, in section \ref{sec:c}, we make the conclusions.

\section{ Model and Molecular Dynamics Simulation \label{sec:mms}}

In this paper, the time evolution of a random packing of 2D particles was simulated by using the MD method. The generalization to the three-dimensional case is straightforward. We consider both the translational and the rotational movement of each particle. The equations of motion of an arbitrary particle $i$ of mass $m_{i}$ and radius $R_{i}$ are:

\begin{figure*}[!t]
 \centering
 \includegraphics[scale=0.90]{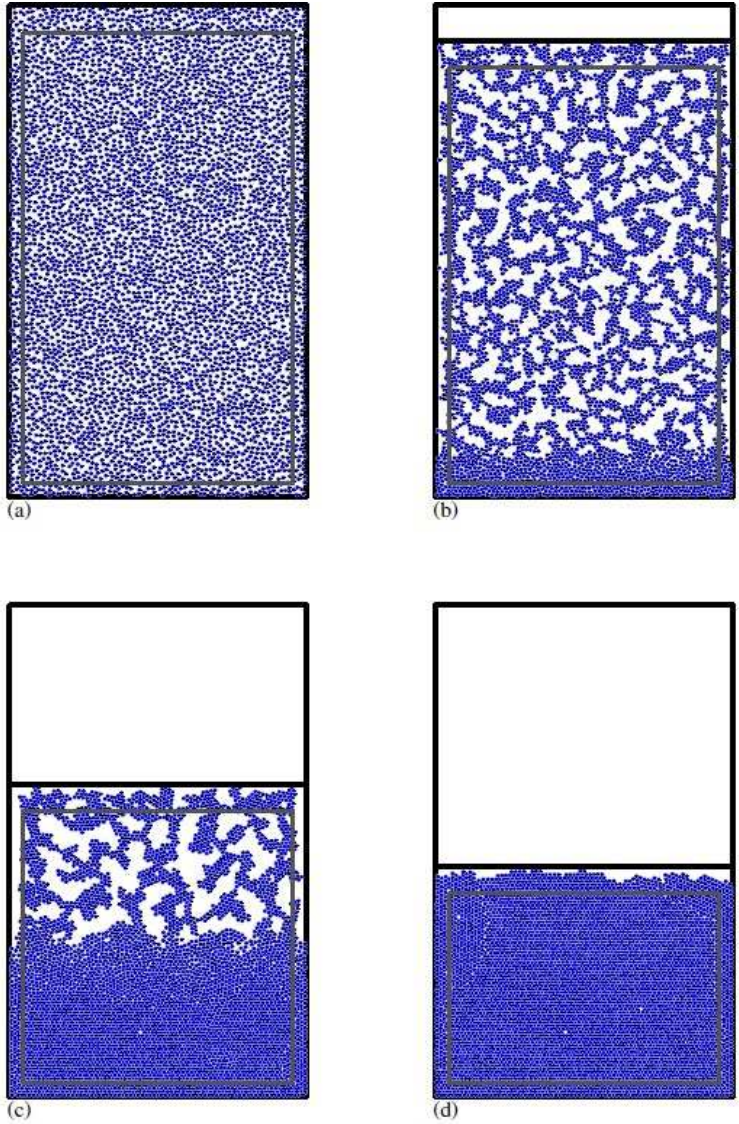}
 \caption{Snapshots of the packing process of monosized particles ($R=2.0\, \mu m$) inside a $300 \, \mu m \times 500 \, \mu m$ box at the instants $t=0.0 \, ms$ (a), $t=1.0 \, ms$ (b), $t=3.0 \, ms$ (c) and $t=7.0 \, ms$ (d). The parameters used in this simulation are given by Table \ref{table:01} for an interaction strength of $\varepsilon_{min}=-10.0\, \mu J$.}\label{fig:02}
\end{figure*}

\begin{figure*}[!t]
 \centering
 \includegraphics[scale=0.90]{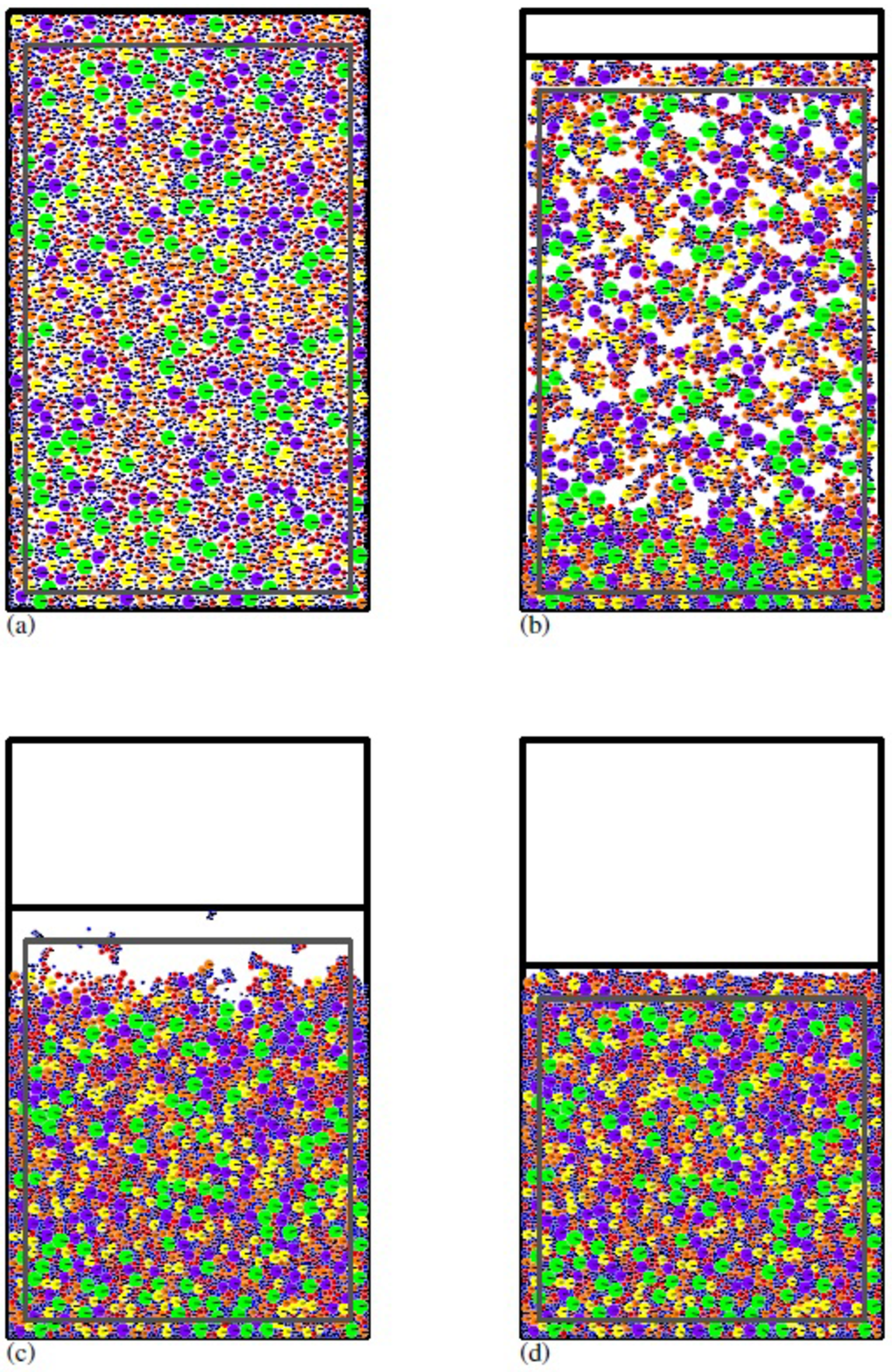}
\caption{Snapshots of the packing process of polydispersive particles (radii ranging from $1.0 \, \mu m$ to $7.0 \, \mu m$) inside a $300 \, \mu m \times 500 \, \mu m$ box at the instants $t=0.0 \, ms$ (a), $t=1.0 \, ms$ (b), $t=3.0 \, ms$ (c) and $t=7.0 \, ms$ (d). The parameters used in this simulation are given by Table \ref{table:01} for an interaction strength of $\varepsilon_{min}=-10.0\, \mu J$.}\label{fig:03}
\end{figure*}

\begin{figure*}[!t]
\centering
\begin{minipage}[t]{1.0\linewidth}
\subfigure[Soft particles]{\label{fig:04a}\includegraphics[scale=0.30, angle=0]{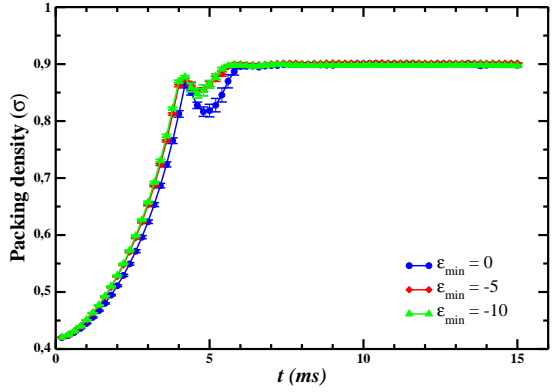}}
\hspace{1.0cm}
\subfigure[Hard particles]{\label{fig:04b}\includegraphics[scale=0.30, angle=0]{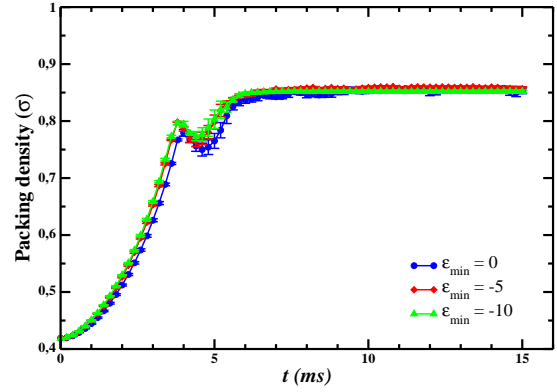}}
\caption{Plot of the packing density of $2.0 \, \mu m$ particles as a function of time for different values of the interaction strength. (a) Soft particles (particles with parameters given by Table \ref{table:01}). (b) Hard particles (same as in (a) but $k_{n}=1.0\times10^{12} \, N/m$). }\label{fig:04}
\vspace{0.50cm}
\end{minipage}
\end{figure*}

\begin{equation} \label{eq:1}
m_{i}\dfrac{d^{2}\mathbf{r}_{i}}{dt^{2}} = \mathbf{F}_{i}
\end{equation}

\begin{equation} \label{eq:2}
 I_{i}\dfrac{d\pmb{\omega}_{i}}{dt}=\mathbf{M}_{i},
\end{equation}
where $\mathbf{r}_{i}$ is the position, $\pmb{\omega}_{i}$ is the angular velocity and $I_{i}=1/2m_{i}R_{i}^{2}$ is the moment of inertia of the particle. While $\mathbf{F}_{i}$ and $\mathbf{M}_{i}$ are the resultant force and torque acting on the particle, respectively. The resultant force $\mathbf{F}_{i}$ is given by

\begin{equation} \label{eq:3}
 \mathbf{F}_{i} = \sum_{j}(\mathbf{F}_{ij}^{n}+\mathbf{F}_{ij}^{t}+\mathbf{F}_{ij}^{lj})+m_{i}\mathbf{g}-\mathbf{f}_{i}^{b},
\end{equation}
where $\mathbf{F}_{ij}^{n}$ is the normal viscoelastic force, $\mathbf{F}_{ij}^{t}$ is the tangential friction force, $\mathbf{F}_{ij}^{lj}$ is the Lennard-Jones force between the particles $i$ and $j$, $\mathbf{f}_{i}^{b}$ is the background friction force, and $\mathbf{g}$ is the gravity acceleration.

The normal viscoelastic force $\mathbf{F}_{ij}^{n}$ is written as
\begin{equation} \label{eq:4}
 \mathbf{F}_{ij}^{n}=k_{n}\xi(t)\medspace \mathbf{\hat{n}}_{ij}-\gamma_{n} \mathbf{v}_{n},
\end{equation}
 where $k_{n}$ is the contact normal stiffness, $\xi(t)$ is the deformation expressed by
 \begin{equation}\label{555}
    \xi(t)=(R_{i}+R_{j})-(\mathbf{r}_{i}(t)-\mathbf{r}_{j}(t))\cdot\mathbf{\hat{n}}_{ij},
 \end{equation}
 being  $\mathbf{\hat{n}}_{ij}$ the unit vector whose direction is from the center of particle $i$ to that of particle $j$, $\gamma_{n}$ is the viscous normal damping and $\mathbf{v}_{n}$ is the relative velocity in the normal direction which is calculated by
\begin{equation}\label{55}
 \mathbf{v}_{n}=[(\mathbf{v}_{i}-\mathbf{v}_{j})\cdot\mathbf{\hat{n}}_{ij}]\medspace \mathbf{\hat{n}}_{ij}.
\end{equation}

Note that the first term in Eq.~(\ref{eq:4}) consists of an elastic, conservative term due to the deformation of the particles and has the same linear dependence on $\xi$ than the Hertzian contact force between cylinders with parallel axes~\cite{popov2010}. The second viscous term is due to dissipation of energy suffered by the particles during the collision, which depends on the deformation rate $\dot{\xi}$~\cite{nesterenko2013,brilliantov1996}.

The tangential friction force is coupled to the normal force via Coulomb's law, i.e, $F_{ij}^{t}\leq\mu_{s}F_{ij}^{n}$. For the dynamic case, one has dynamic (sliding) friction  given by $F_{ij}^{t}=\mu_{d}F_{ij}^{n}$. In general, the dynamic and static friction coefficients follow the relation $\mu_{d}\leq \mu_{s}$. For the static situation, we can define a tangential elastic spring of length $\delta$ in order to allow for a restoring force due to the Coulomb friction between the colliding particles. To check the changes of $\delta$, we first compute the tangential test-force $\mathbf{f}_{t}$ as
\begin{equation}\label{eq:5}
 \mathbf{f}_{t}=-k_{t}\delta(t) \medspace \mathbf{t}_{ij}-\gamma_{t}\mathbf{v}_{t}.
\end{equation}
In Eq.~(\ref{eq:5}), $k_{t}$ is the contact tangential stiffness, $\mathbf{t}_{ij}$ is the unit vector along the tangential direction, $\mathbf{v}_{t}$ is the relative velocity in the tangential direction that is given by
\begin{equation}\label{eq:6}
 \mathbf{v}_{t}=[(\mathbf{v}_{i}-\mathbf{v}_{j})\cdot\mathbf{t}_{ij}]\medspace\mathbf{t}_{ij}+(R_{i}\mathbf{\hat{n}}_{ij}
 \times\pmb{\omega}_{i}+R_{j}\mathbf{\hat{n}}_{ij}\times \pmb{\omega}_{j}).
\end{equation}
From Eqs.~(\ref{eq:5}) and (\ref{eq:6}) and taking into account the Coulomb's law, the tangential friction force $\mathbf{F}_{ij}^{t}$ is calculated by

\begin{equation} \label{eq:7}
\mathbf{F}_{ij}^{t}=\begin{cases}
 \mathbf{f}_{t},&\text{if $|\mathbf{f}_{t}|\leq \mu_{s}F_{ij}^{n}$,} \\
 -\mu_{d}F_{ij}^{n}\mathbf{t}_{ij},&\text{if $|\mathbf{f}_{t}|> \mu_{s}F_{ij}^{n}$}.
 \end{cases}
\end{equation}

The long-range forces between the particles $i$ and $j$ are calculated only in the region $r>R_{i}+R_{j}$ by using the Lennard-Jones potential, which is defined as
\begin{equation} \label{eq:8}
 \Phi_{ij}^{lj}=4\varepsilon_{min} \Bigl[\Bigl(\dfrac{\sigma}{r}\Bigr)^{6}-\Bigl(\dfrac{\sigma}{r}\Bigr)^{12}\Bigr],
\end{equation}
where $\varepsilon_{min}<0$, is the minimum potential energy which governs the strength of the interaction and $\sigma=2^{-1/6}(R_{i}+R_{j})$ defines the hard core of the potential. For the region $r\leq R_{i}+R_{j}$, both the normal viscoelastic (Eq.~\ref{eq:4}) and tangential friction (Eq.~\ref{eq:7}) forces take place on colliding particles. The Lennard-Jones force between the particles $i$ and $j$ can be evaluated as

\begin{equation} \label{eq:9}
 \mathbf{F}_{ij}^{lj}=-\pmb{\bigtriangledown}\Phi_{ij}^{lj}=\dfrac{24\varepsilon_{min}}{\sigma}\Bigl[\Bigl(\dfrac{\sigma}{r}\Bigr)^{7}-2\Bigl(\dfrac{\sigma}{r}\Bigr)^{13}\Bigr]\mathbf{\hat{n}}_{ij}.
\end{equation}
In Fig.~\ref{fig:01} are plotted both the potential $\Phi_{ij}^{lj}/\epsilon_{min}$ and force $F_{ij}^{lj}/\epsilon_{min}$ as functions of the relative distance $r/(R_{i}+R_{j})$ between a pair of particles of radii $R_{i}$ and $R_{j}$. In order to save computational effort, we have used a cutoff at $r=3.0(R_{i}+R_{j})$ to compute the Lennard-Jones force.

The torque $\mathbf{M}_{i}$ acting on a given particle $i$ is written as
\begin{equation} \label{eq:10}
 \mathbf{M}_{i}=\sum_{j} \mathbf{\hat{n}}_{ij}R_{i} \times \mathbf{F}_{ij}^{t}.
\end{equation}
With the purpose of reducing computation time, we do not consider the rolling frictional effect in Eq.~(\ref{eq:10}). However, due to the long-wavelength cooperative modes~\cite{luding94} present in the dynamics of many colliding particles, the dissipation forces are rather inefficient to obtain a rapid relaxation and equilibration of the system. Therefore, we have also added a background friction force $\mathbf{f}_{i}^{b}=-\gamma_{b}\mathbf{v}_{i}$ in Eq.~(\ref{eq:3}) in order to decrease both the relaxation and the equilibration time in the simulations. Of course, all simulations were monitored in order to prevent possible unrealistic effects.

A symplectic leapfrog scheme~\cite{rapaport95} was used to integrate numerically the Eqs.~(\ref{eq:1}) and (\ref{eq:2}). The positions and velocities are calculated as
\begin{equation}\label{eq:11aa}
     \mathbf{r}_{i}(t)=\mathbf{r}_{i}(t-\Delta t)+\mathbf{v}_{i}(t-\Delta t/2)\medspace \Delta t,
\end{equation}
\begin{equation}\label{eq:11a}
    \mathbf{v}_{i}(t+\Delta t/2)=\mathbf{v}_{i}(t-\Delta t/2)+\dfrac{\mathbf{F}_{i}(t)}{m_{i}} \Delta t,
\end{equation}
\begin{equation}\label{eq:11bb}
    \theta_{i}(t)=\theta_{i}(t-\Delta t)+\omega_{i}(t-\Delta t/2)\medspace \Delta t,
\end{equation}
\begin{equation}\label{eq:11b}
    \omega_{i}(t+\Delta t/2)=\omega_{i}(t-\Delta t/2)+\dfrac{M_{i}(t)}{I_{i}} \Delta t.
\end{equation}

In Eqs.~\ref{eq:11a} and \ref{eq:11b}, $\mathbf{F}_{i}$ and $M_{i}$ are calculated, respectively, by Eqs.~(\ref{eq:3}) and (\ref{eq:10}) at each time-step $\Delta t$. In this scheme, the stability is achieved when $\Delta t \leq 1/\Omega$.  Here $\Omega$ can be understood as the normal frequency of response during the contact between two particles, which can be expressed as~\cite{luding98}
\begin{equation} \label{eq:12}
 \Omega=\sqrt{k_{n}/m_{ij}-(\gamma_{eff}/2m_{ij})^{2}},
\end{equation}
being $m_{ij}=m_{i}m_{j}/(m_{i}+m_{j})$ the two-body reduced mass and $\gamma_{eff}$ the effective damping over the particles. The contact time between two particles is roughly given by $T_{c}=\pi/\Omega$.

The physical parameters used in the simulations were obtained by trial tests and are given in Table \ref{table:01}. In order to avoid the complicating effects of the pouring rate, the particles were suspended along the box at the beginning of the simulation. Owing to frictional forces, stable simulations were already achieved by taking a time-step $\Delta t = 10^{-6} s$. The average CPU time to update the state of one particle was approximately $0.16 \thinspace ms$ on one 3.70 GHz Intel Xeon microprocessor.

\begin{figure*}[!t]
\begin{minipage} [t]{0.49\linewidth}
\includegraphics*[scale=0.30,angle=0]{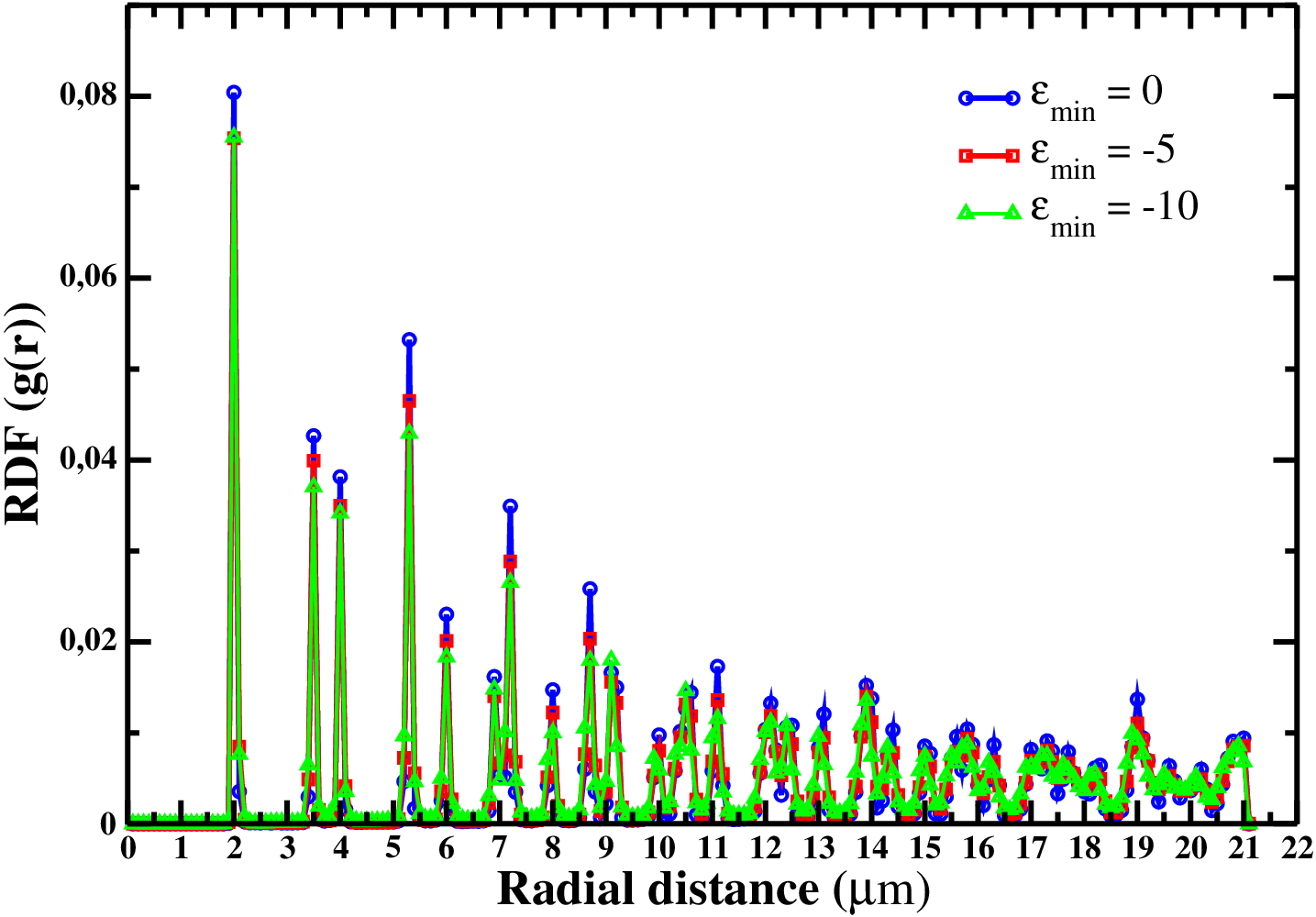}
\caption{RDFs versus the radial distance of the RCPS of $1.0 \,\mu m$ soft particles for different $\varepsilon_{min}$ values.}\label{fig:07}
\end{minipage}\hfill
\begin{minipage}[t]{0.49\linewidth}
\includegraphics*[scale=0.30,angle=0]{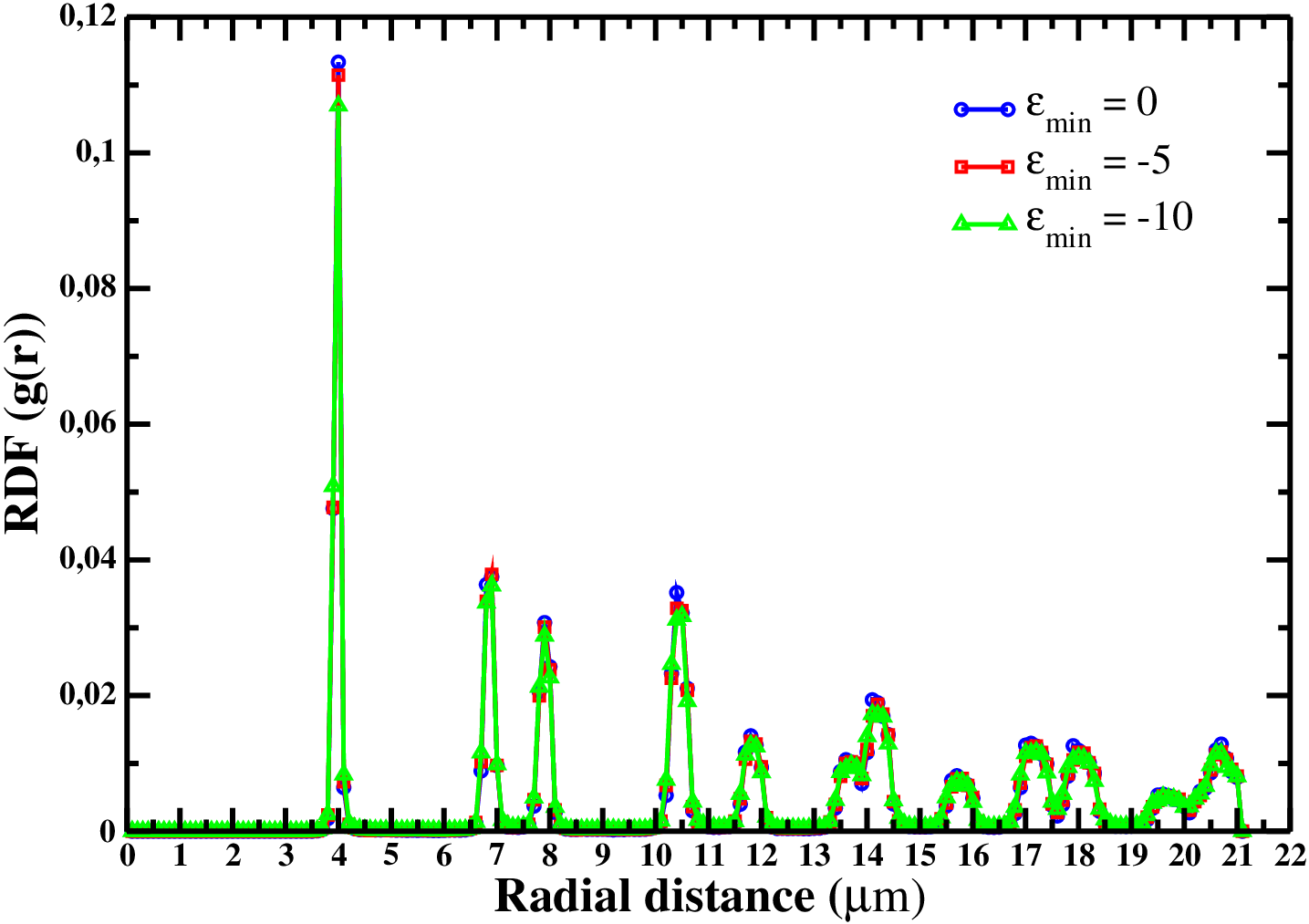}
\caption{RDFs versus the radial distance of the RCPS of $2.0 \,\mu m$ soft particles for different $\varepsilon_{min}$ values.}\label{fig:08}
\vspace{0.50cm}
\end{minipage}
\end{figure*}

\begin{figure*}[!t]
\begin{minipage}[t]{0.49\linewidth}
\includegraphics*[scale=0.30,angle=0]{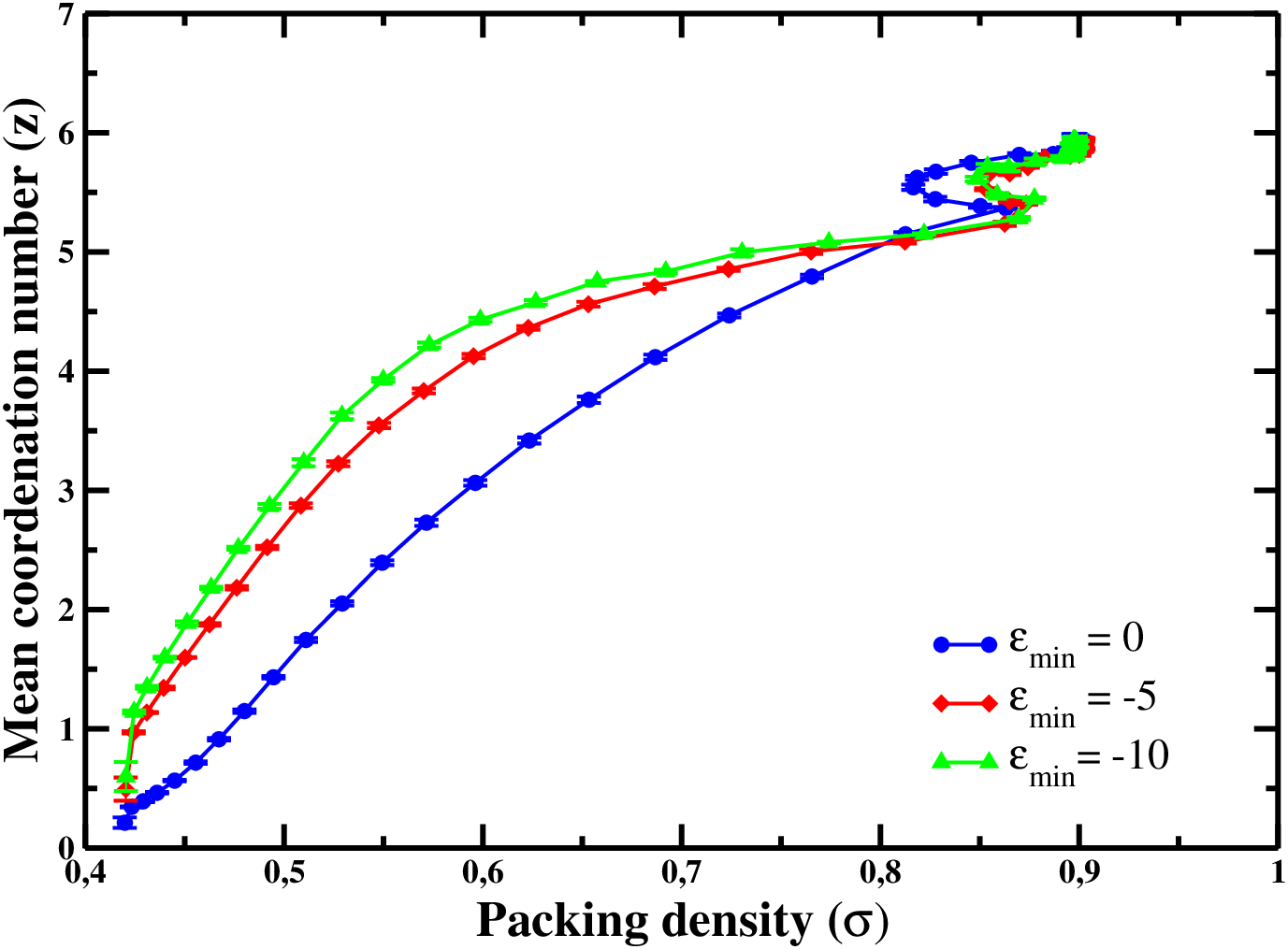}
\caption{Plot of the mean coordination number $z$ of $2.0 \, \mu m$ soft particles against the packing density for different $\varepsilon_{min}$ values.}\label{fig:05}
\end{minipage}\hfill
\begin{minipage} [t]{0.49\linewidth}
\includegraphics*[scale=0.30,angle=0]{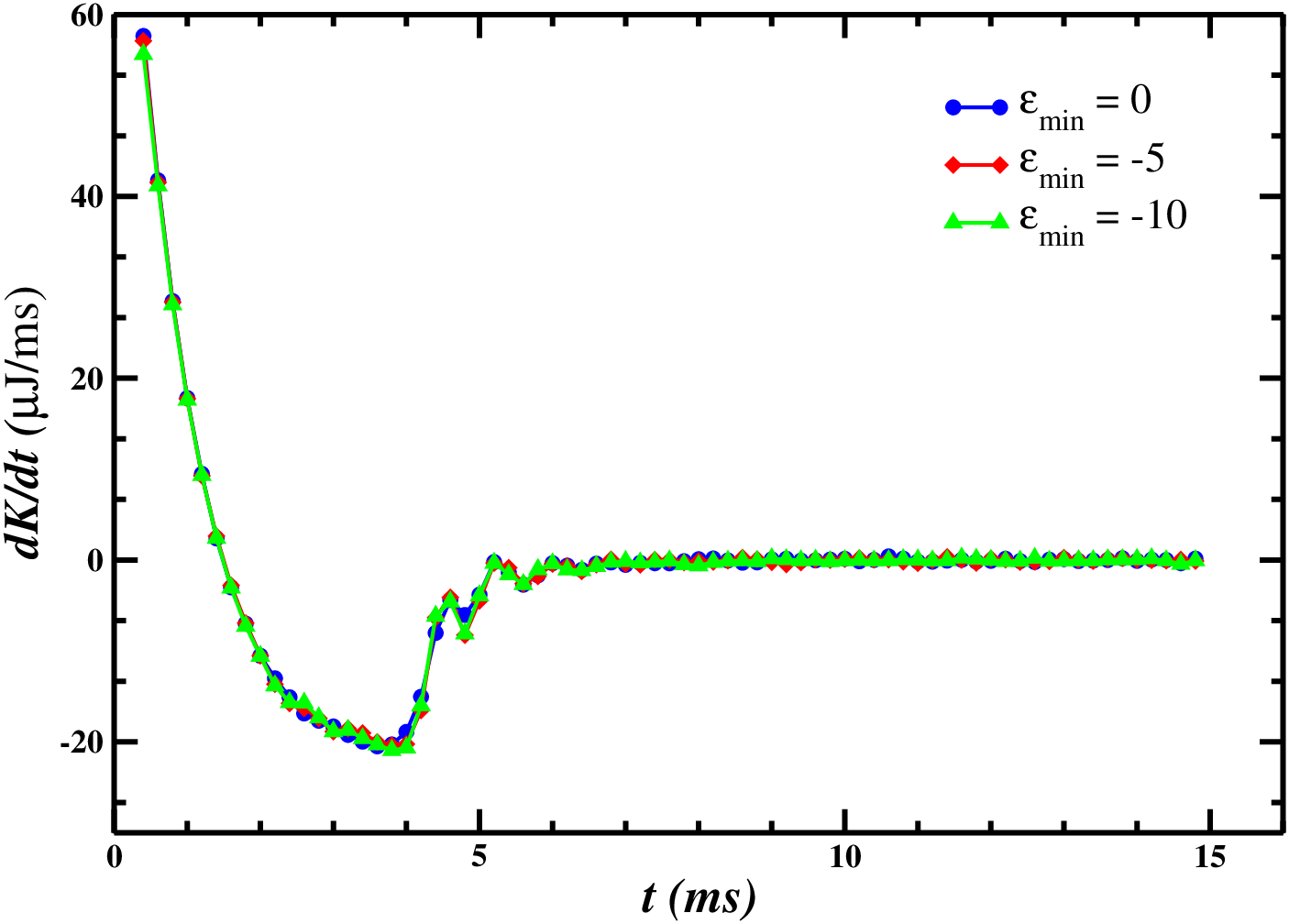}
\caption{Plot of the time derivative of the kinetic energy of $2.0 \, \mu m$ soft particles as a function of time for different $\varepsilon_{min}$ values. The relaxation time is about $6.0 \, ms$.}\label{fig:06}
\vspace{0.50cm}
\end{minipage}
\end{figure*}

\begin{figure*}[!t]
\centering
\begin{minipage}[t]{1.0\linewidth}
\subfigure[Soft particles]{\label{fig:09a}\includegraphics[scale=0.30, angle=0]{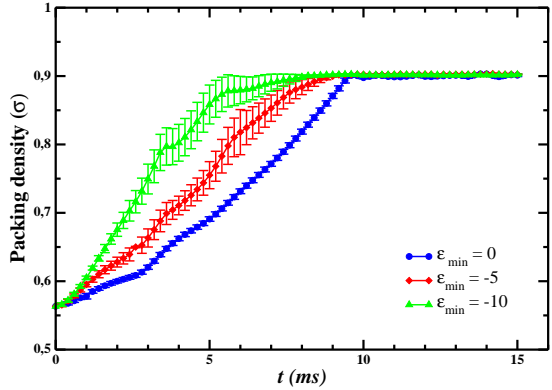}}
\hspace{1.0cm}
\subfigure[Hard particles]{\label{fig:09b}\includegraphics[scale=0.30, angle=0]{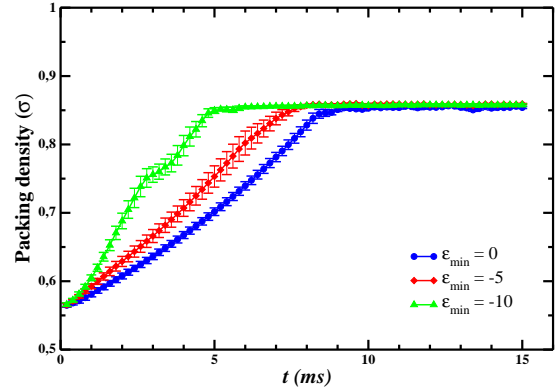}}
\caption{Plot of the packing density of polydispersive particles with random size distribution as a function of time for different values of the interaction strength. (a) Soft particles (particles with parameters given by Table \ref{table:01}). (b) Hard particles (same as in (a) but with $k_{n}=1.0\times10^{12} \, N/m$). }\label{fig:09}
\vspace{0.50cm}
\end{minipage}
\end{figure*}

\begin{figure*}[!t]
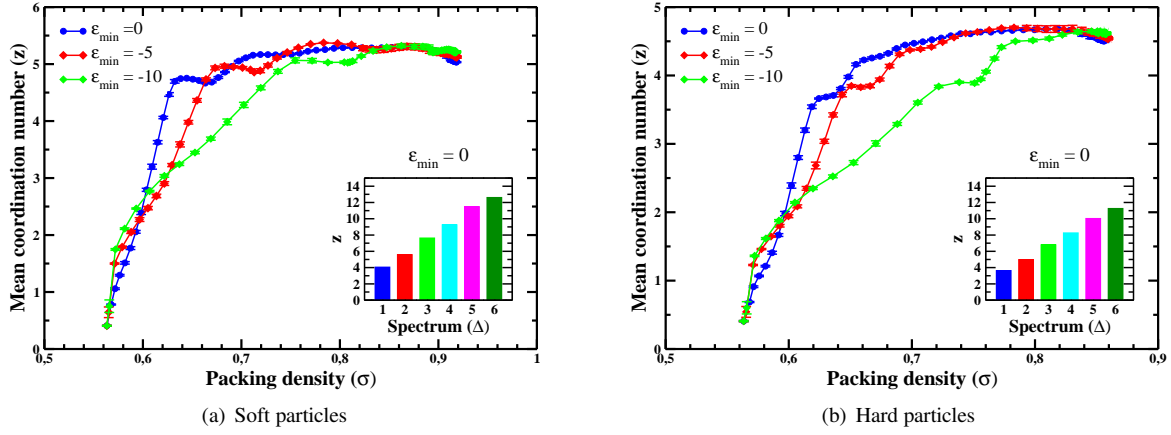

\centering
\begin{minipage}[t]{1.0\linewidth}
\subfigure[Soft particles]{\label{fig:10a}\includegraphics[scale=0.30, angle=0]{cnspoli.eps}}
\hspace{1.0cm}
\subfigure[Hard particles]{\label{fig:10b}\includegraphics[scale=0.30, angle=0]{cnspoli16.eps}}
\caption{Plot of the mean coordination number ($z$) of polydispersive particles with random size distribution against the packing density for different $\varepsilon_{min}$ values. (a) Soft particles (particles with parameters given by Table \ref{table:01}). (b) Hard particles (same as in (a) but $k_{n}=1.0\times10^{12} \, N/m$). Insets give the mean coordination number in each spectrum of the particles that compound the RCPS for the case $\varepsilon_{min}=0$.}\label{fig:10}
\vspace{0.50cm}
\end{minipage}
\end{figure*}

\begin{figure*}[!t]
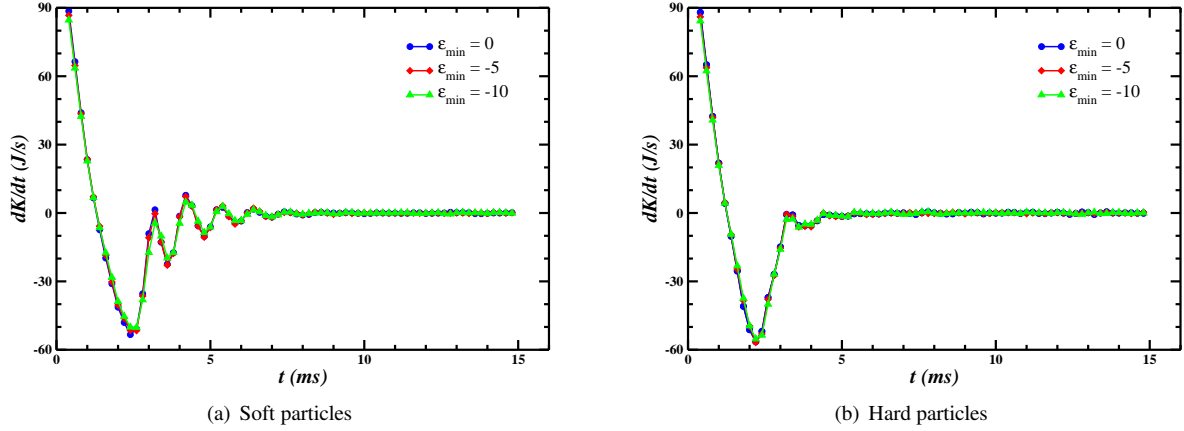

\centering
\begin{minipage}[t]{1.0\linewidth}
\subfigure[Soft particles]{\label{fig:11a}\includegraphics[scale=0.30, angle=0]{energypoli.eps}}
\hspace{1.0cm}
\subfigure[Hard particles]{\label{fig:11b}\includegraphics[scale=0.30, angle=0]{energypoli16.eps}}
\caption{Plot of the time derivative of the kinetic energy of polydispersive particles with random size distribution as a function of time for different $\varepsilon_{min}$ values. (a) Soft particles (particles with parameters given by Table \ref{table:01}). (b) Hard particles (same as in (a) but $k_{n}=1.0\times10^{12} \, N/m$).}\label{fig:11}
\vspace{0.50cm}
\end{minipage}
\end{figure*}

\begin{figure*}[!t]
\centering
\begin{minipage}[t]{1.0\linewidth}
\subfigure[Soft particles]{\label{fig:12a}\includegraphics[scale=0.30, angle=0]{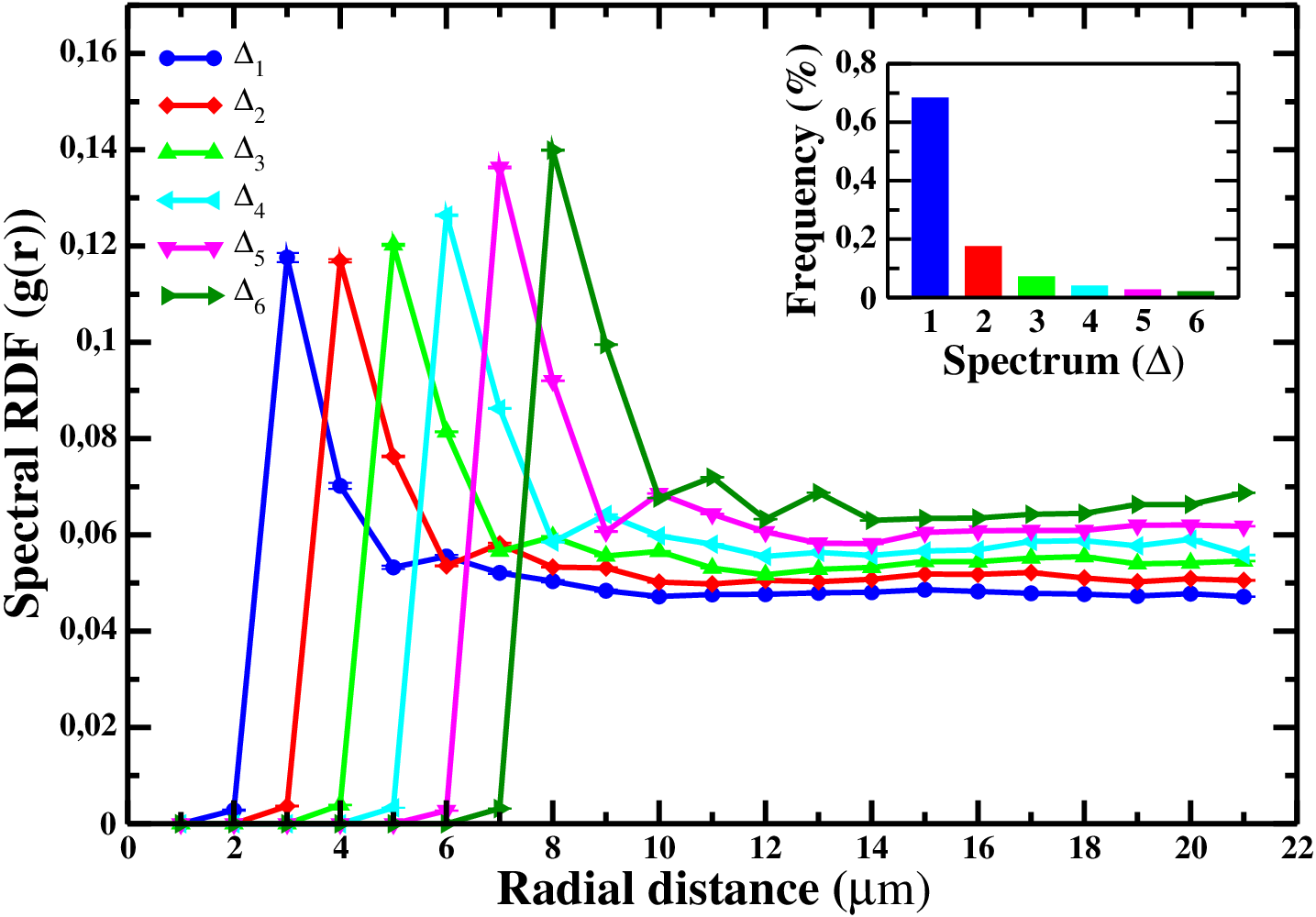}}
\hspace{1.0cm}
\subfigure[Hard particles]{\label{fig:12b}\includegraphics[scale=0.30, angle=0]{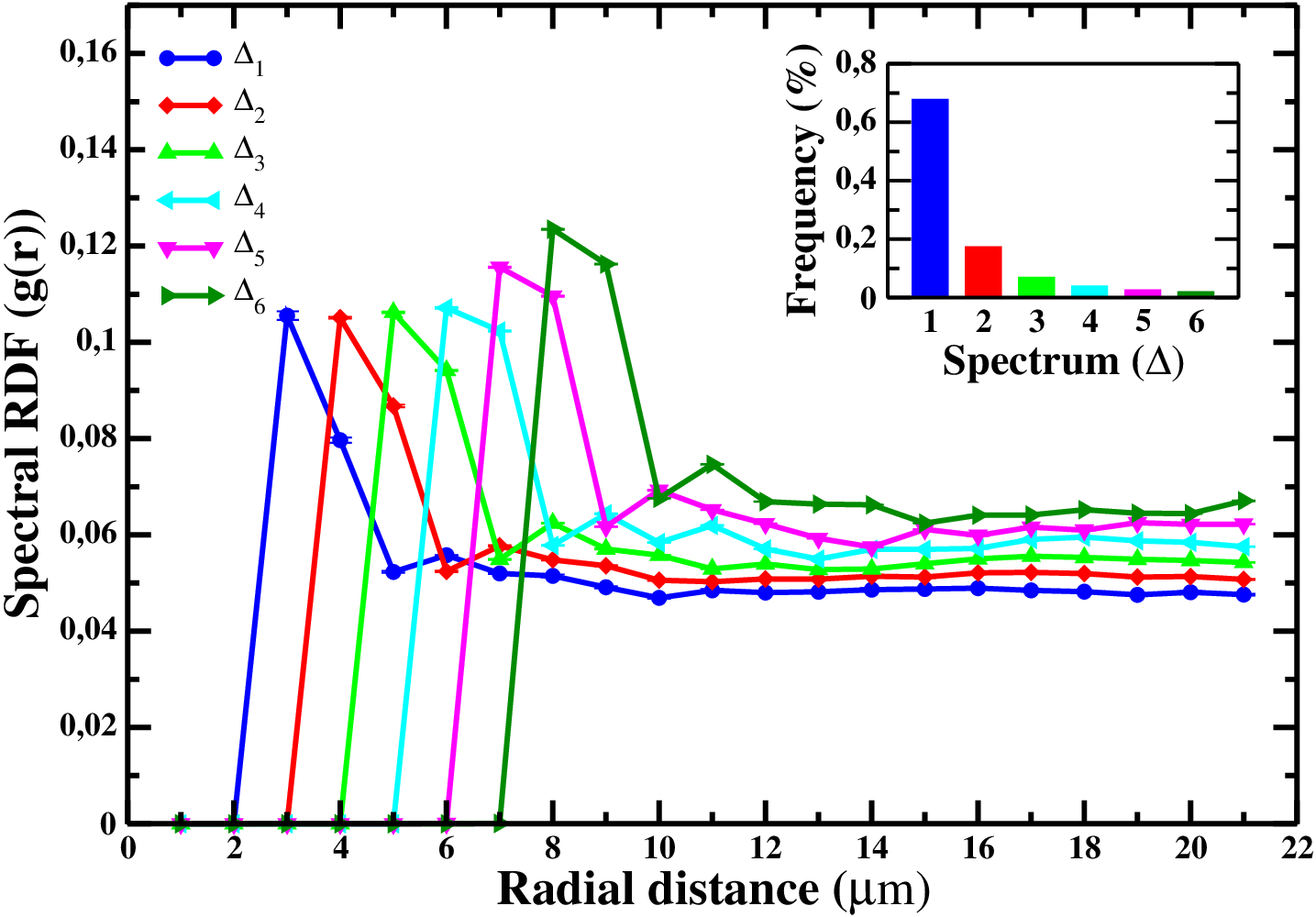}}
\caption{ Spectral RDFs as functions of the radial distance of the  RCPS formed by polydispersive particles considering $\varepsilon_{min}=0$. Fig.~\ref{fig:12a} for soft particles and Fig.~\ref{fig:12b} for hard ones. Insets give the frequency of the particles for each spectrum.}\label{fig:12}
\vspace{0.50cm}
\end{minipage}
\end{figure*}

\begin{figure*}[!t]
\centering
\begin{minipage}[t]{1.0\linewidth}
\subfigure[Soft particles]{\label{fig:13a}\includegraphics[scale=0.30, angle=0]{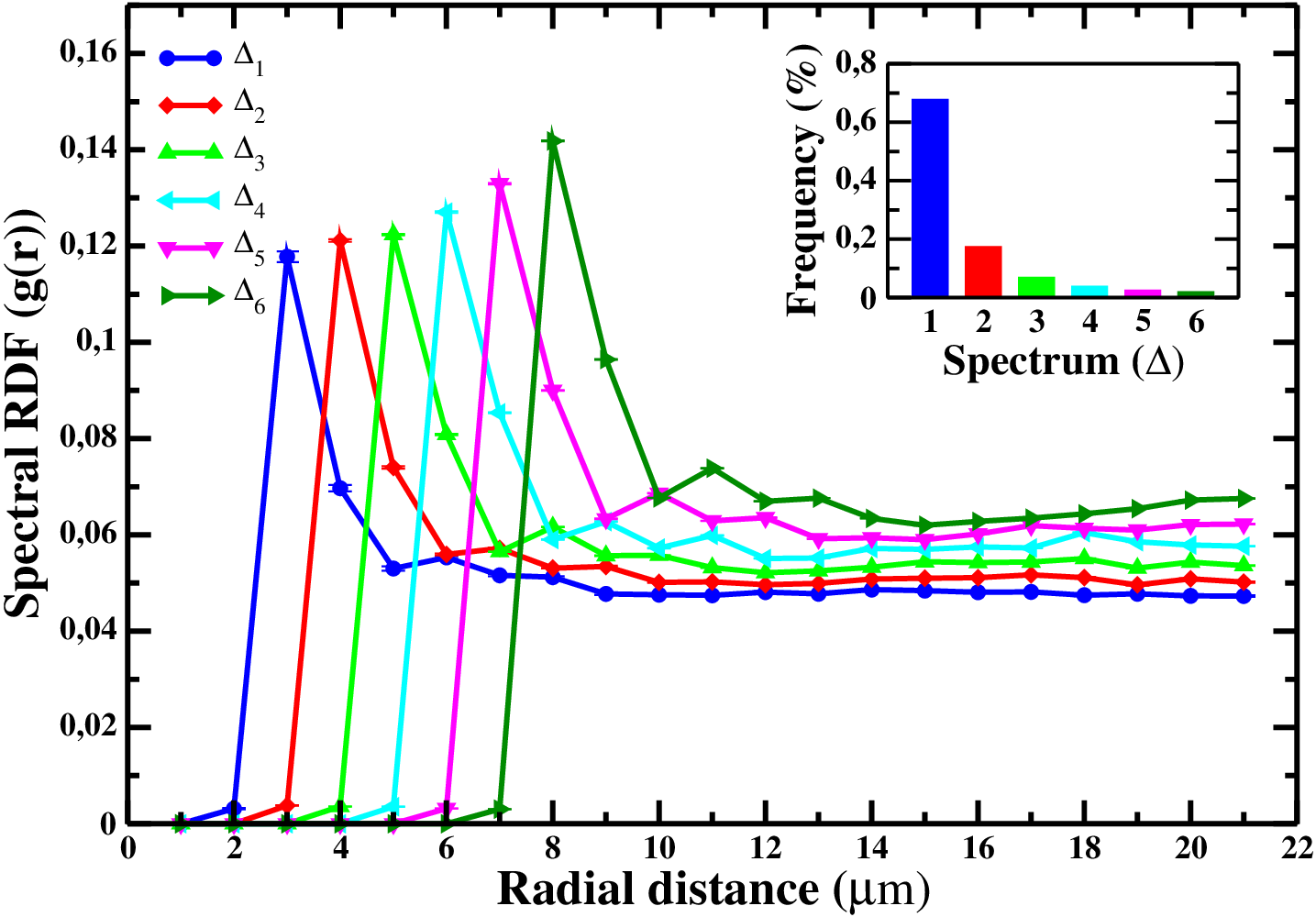}}
\hspace{1.0cm}
\subfigure[Hard particles]{\label{fig:13b}\includegraphics[scale=0.30, angle=0]{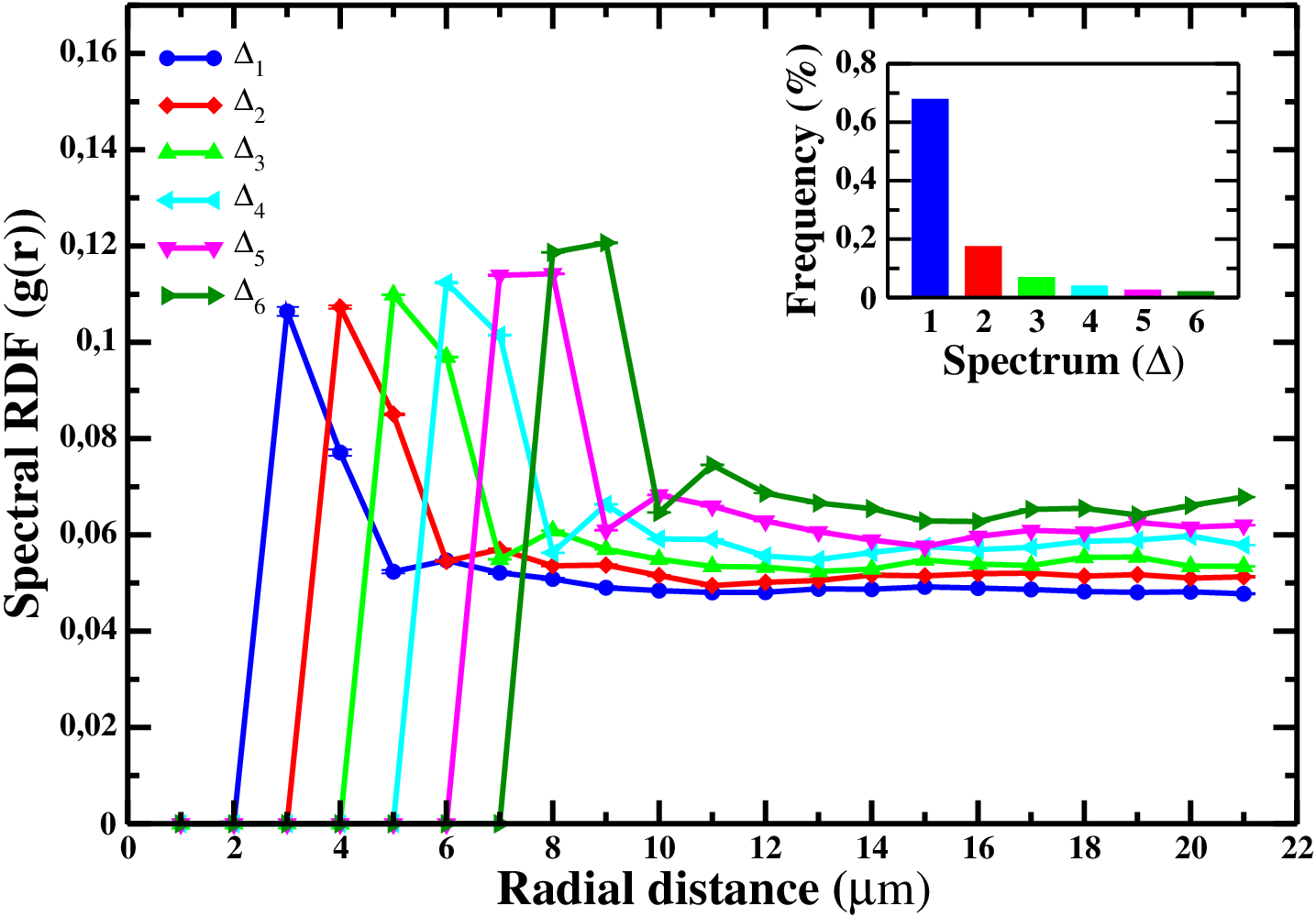}}
\caption{ Spectral RDFs as functions of the radial distance of the  RCPS formed by polydispersive particles considering $\varepsilon_{min}=-5.0 \, \mu J$. Fig.~\ref{fig:13a} for soft particles and Fig.~\ref{fig:13b} for hard ones. Insets give the frequency of the particles for each spectrum.}\label{fig:13}
\vspace{0.50cm}
\end{minipage}
\end{figure*}

\begin{figure*}[!t]
\centering
\begin{minipage}[t]{1.0\linewidth}
\subfigure[Soft particles]{\label{fig:14a}\includegraphics[scale=0.30, angle=0]{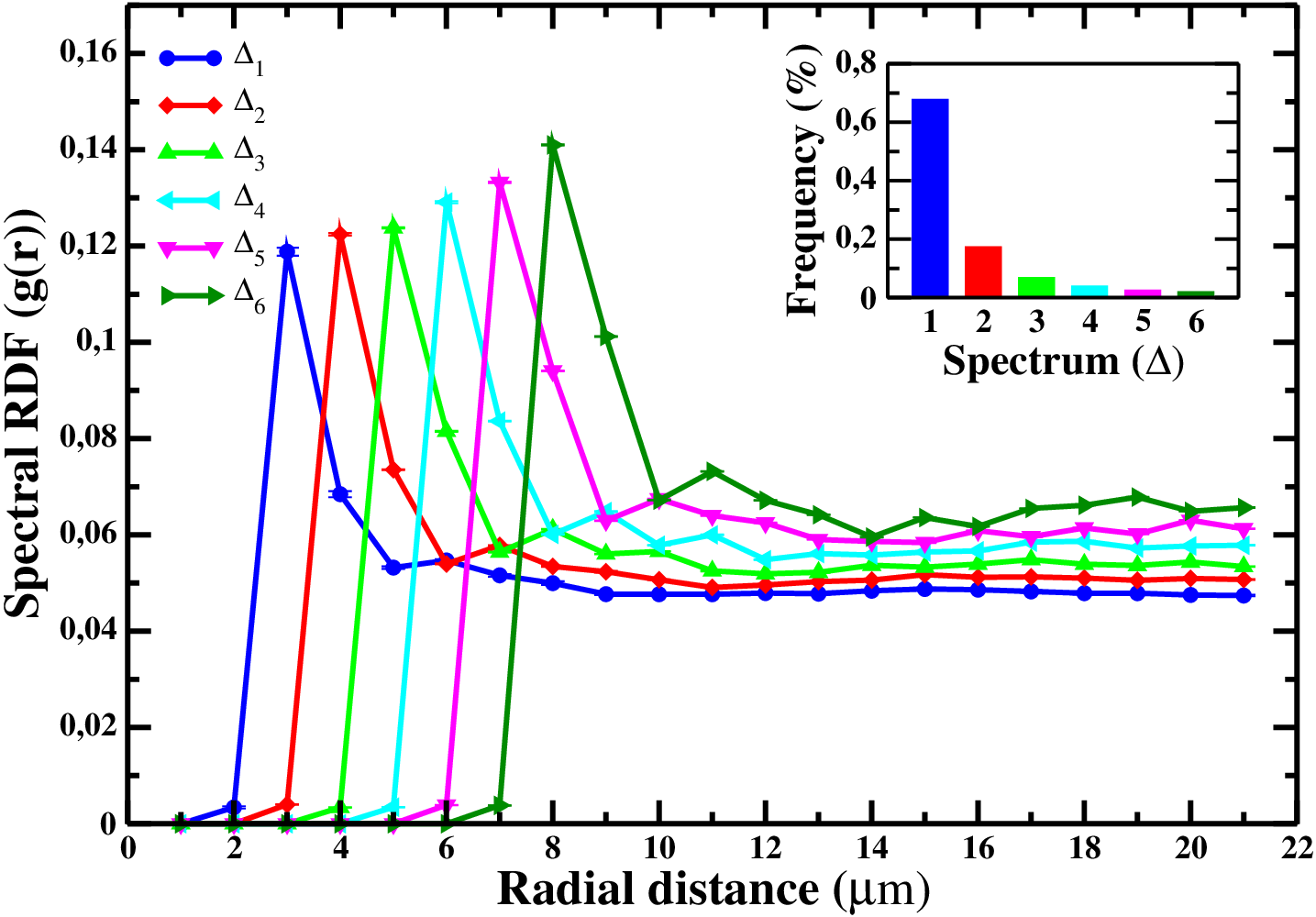}}
\hspace{1.0cm}
\subfigure[Hard particles]{\label{fig:14b}\includegraphics[scale=0.30, angle=0]{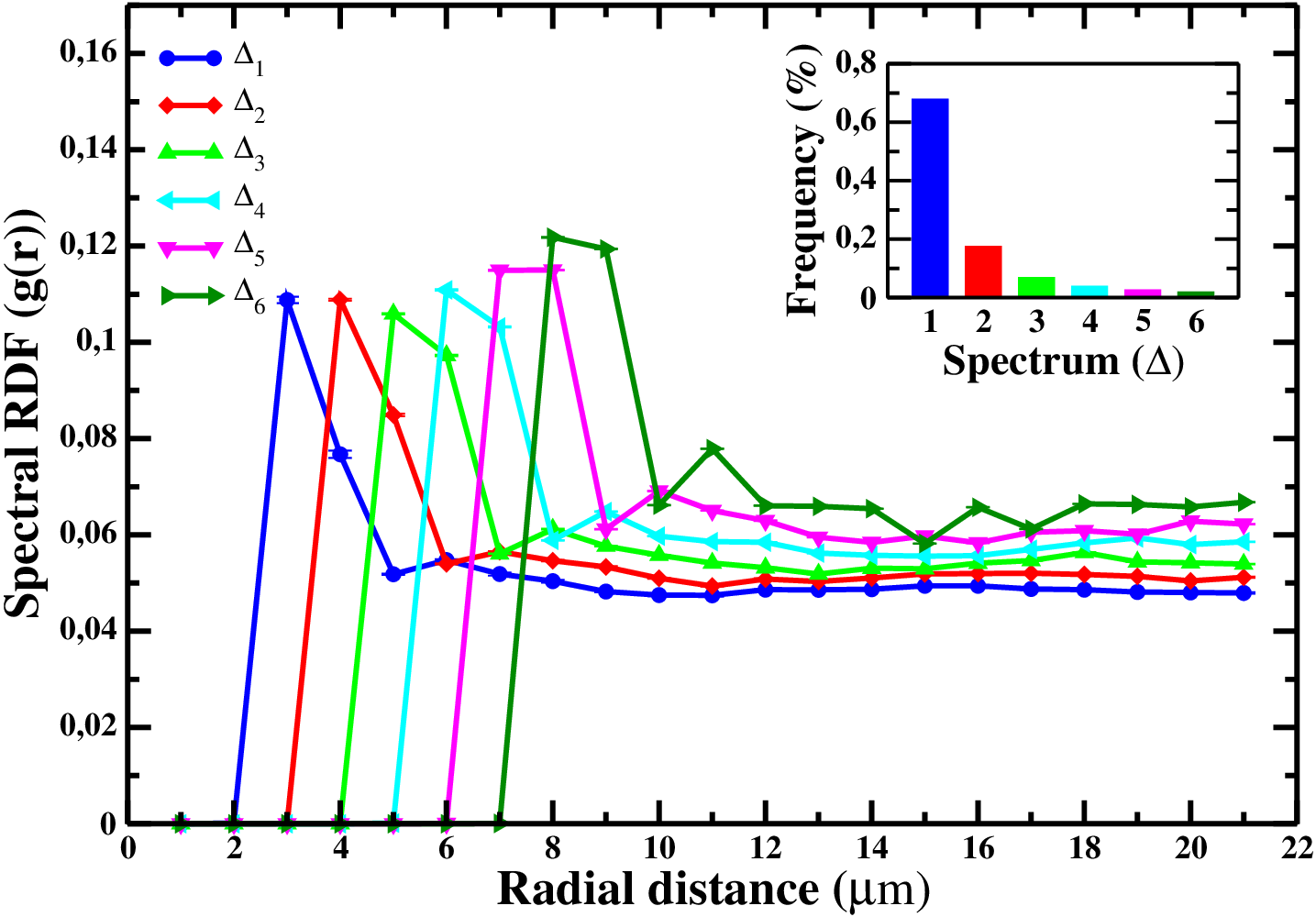}}
\caption{Spectral RDFs as functions of the radial distance of the  RCPS formed by polydispersive particles considering $\varepsilon_{min}=-10.0 \, \mu J$. Fig.~\ref{fig:14a} for soft particles and Fig.~(\ref{fig:14b}) for hard ones. Insets give the frequency of the particles for each spectrum.}\label{fig:14}
\end{minipage}
\end{figure*}

\section{\label{sec:r} Results and Discussion}

The packing method by pouring of particles employed here is one of the most studied in the literature. Its major advantage is to get high packing densities by means of a simple and natural (gravity action) mechanism. It was already shown that improvements in the compaction of the particles can be accomplished when one shakes the formed aggregate, yielding particle packings with the highest densities. However, a sufficient time is required to achieve the saturation density~\cite{he99}, which depends on the shake amplitude. From previous numerical and experimental works~\cite{Bideau2004}, this saturation density increases as the shake amplitude decreases. Thus, more time and energy are demanded when one introduces vibrating walls in the system. Although these wall effects are not included in this study, the elastic properties of the particles generate transient vibrations during the collisions of these with bottom of the box.

The packing process is depicted in Figs.~\ref{fig:02} and \ref{fig:03} for monosized and polydispersive particles, respectively. Snapshots at the instants $t=0.0 \, ms$, $t=1.0 \, ms$, $t=3.0 \, ms$ and $t=7.0 \, ms$ are shown in these figures. The parameters used in these simulations are given by Table \ref{table:01} for $\varepsilon_{min}=-10.0\, \mu J$. The thick black straight line inside the box indicates the highest positions occupied by the particles and helps to localize the packing in order to calculate the packing density $\sigma$.  In addition, the black straight line running from the center of each particle to a fixed point on its circumference gives an indication of the angular position of the particle at each instant. The thick gray lines in these figures represent an imaginary box used to eliminate wall effects in the calculations of the mean coordination number $z$. For an interaction strength $\varepsilon_{min}=-10.0 \, \mu J$, one can notice the formation of a large particle cluster during the packing process. This large cluster is especially important in reducing the particles rearrangement during the formation of the RCPS. In addition, the formed RCPS has an irregular superior layer as seen in Figs.\, 2(d) and 3(d).  We perform statistical calculations of the different quantities studied as the system evolves over time such as packing density, mean coordination number, time derivative of the kinetic energy and RDF.  To determine the average value of these quantities and estimate the statistical error, we average over a number of independent realizations of the packing process. For the monodispersive case, we average over six independent realizations, whereas for the polydispersive case, ten independent realizations are used.  Here, we consider two type of particles: soft particles (so-called particles with $k_{n}=1.0 \times 10^{11}\, N/m$) and hard particles ($k_{n}=1.0 \times 10^{12}\, N/m$).

\subsection{Monosized particles}

We study the packing process with $2.0 \,\mu m$ particles for three different values of the interaction strength, namely, $\varepsilon_{min}=0.0$ (absence of long-range forces), $-5.0 \, \mu J$  and $-10.0 \,\mu J$. Fig.~\ref{fig:04} shows the packing density as a function of time for different $\varepsilon_{min}$ values for both soft (Fig.~\ref{fig:04a}) and hard particles (Fig.~\ref{fig:04b}). From Fig.~\ref{fig:04}, it is seen that the main effect of the long-range forces is sightly increase the packing density during the fall of the particles inside the box. The local minimum in the packing density around of $5.0\, ms$ is due to the first bouncing of the particles when they hit the bottom of the box. The initial relative packing density was 0.4198. For soft particles, the packing density obtained after a time of $15.0\, ms$ was $0.888 \pm 0.003$ for $\varepsilon_{min}=0.0 $; $0.891 \pm 0.001$ for both $\varepsilon_{min}=-5.0 \, \mu J$ and $\varepsilon_{min}=-10.0 \, \mu J$. For hard particles, the packing density obtained at the same time was $0.853 \pm 0.005$ for $\varepsilon_{min}=0.0 $; $0.857 \pm 0.003$ for $\varepsilon_{min}=-5.0 \, \mu J$; and $0.852 \pm 0.003$ for $\varepsilon_{min}=-10.0 \, \mu J $. These results are higher than the packing density of a dense random packing of hard disks reported in the literature of 0.823~\cite{berryman83, german89}. It is interesting to notice that the density values obtained for soft particles are close to that found for the hexagonal packing arrangement of disks, which has the highest density of all possible plane packings, $\dfrac{\pi}{\sqrt{12}}\simeq 0.9069$~\cite{chau2010}. This is due to the non-zero overlapping among soft particles during the packing process, which was also taken into account in the density calculations. The RDF has been widely used to characterize random structures of spherical particles~\cite{scott62,yen91,jia12} and it can be understood as the probability of finding one particle at a given distance from the center of a reference particle. Here we define RDF as
\begin{equation}\label{eq:13}
 g(r_{i})=\dfrac{n(r_{i})}{2\pi r_{i} \delta r_{i} Z},
\end{equation}
being $n(r_{i})$ the number of centers of particles within the ring $i$ of radius $r_{i}$ and thickness $\delta r_{i}$. In above equation, $Z$ is the normalization factor given by
\begin{equation}\label{eq:14}
 Z=\sum_{i=1}^{N_{r}}\dfrac{n(r_{i})}{2\pi r_{i} \delta r_{i}},
\end{equation}
where $N_{r}$ is the total number of rings considered. For monosized particles, we set $\delta r_{i}=0.01 \, \mu \,m$ and $N_{r}=210$.  Figs.~\ref{fig:07} and \ref{fig:08} show the RDF as a function of the radial distance of the RCPS of $1.0 \,\mu m$ and $2.0 \,\mu m$ particles, respectively. In Fig.~\ref{fig:07}, one can observe that the first three main peaks in the RDF  are localized at the distances $2.0\,\mu m$, $2\sqrt{3}\,\mu m$ and $4.0 \, \mu m$ in accordance with previous works~\cite{yen91,zhang2001,jia12}. There exist slight variations in the height of the peaks of the RDF when one considers different $\varepsilon_{min}$ values. It can be explained as a consequence of the lower particle rearrangement occurred during the formation of the RCPS when long-range forces are present. In Fig.~\ref{fig:08}, we can see the first three main peaks in the RDF localized at the distances $4.0\,\mu m$, $4\sqrt{3}\,\mu m$ and $8.0 \, \mu m$ in similar way as Fig.~\ref{fig:07}. However, for this particle size, it is seen a smaller influence of the long-range forces on the RDF of the formed RCPS. This was already expected, once the intensity of these forces decays the larger the particles are, as pointed out in Refs.~\cite{visser89,yen91,israelachvili92}. Fig.~\ref{fig:05} shows the mean coordination number $z$ against the packing density for different $\varepsilon_{min}$ values. The mean coordination number after a time of $15.0\, ms$ was $5.934 \pm 0.058$ for $\varepsilon_{min}=0.0$; $5.913 \pm 0.041$ for $\varepsilon_{min}=-5.0 \, \mu J$; and $5.936 \pm 0.027$ for $\varepsilon_{min}=-10.0 \, \mu J$. In order to determine the relaxation time in the simulation, the time derivative of the kinetic energy of the system is plotted as a function of time in Fig.~\ref{fig:06}. From this figure, we can see that a time of about $6.0 \, ms$ is sufficient for the system to reach its steady state and form an RCPS.

\subsection{Polydispersive particles}

The packing process of polydispersive particles with random size distribution was studied by considering different $\varepsilon_{min}$ values. The radii of the particles inside the box were assigned at random and ranged from $1.0 \, \mu m$ to $7.0 \, \mu m$. For a further analysis, the effects on the different studied quantities were also compared when one considers different $k_{n}$ values. Figs.~\ref{fig:09a} and \ref{fig:09b} display the packing density as a function of time for different $\varepsilon_{min}$ values considering, respectively, soft and hard particles. The initial packing density was about 0.563 in all cases studied. For the polydispersive case, the influence of the long-range forces over the packing process is clearer than for the monodispersive case. Moreover, the RCPS formed by hard particles has a lower packing density than that one formed by soft particles. This happens because the hard particles suffer more elastic collisions than the soft ones, conserving a bigger amount of kinetic
energy and therefore having a higher vibrating amplitude. For soft particles, we obtain an RCPS density of $0.901 \pm 0.002$ for $\varepsilon_{min}=0.0$ and $0.903 \pm 0.001$ for both $\varepsilon_{min}=-5.0\, \mu J$ and $\varepsilon_{min}=-10.0\, \mu J$. While, for harder particles, we obtain an RCPS density of $0.854 \pm 0.001$ for $\varepsilon_{min}=0.0$ and $0.858 \pm 0.002$ for both $\varepsilon_{min}=-5.0\, \mu J$ and $\varepsilon_{min}=-10.0\, \mu J$. The high packing densities obtained in both cases are mainly due to the presence of smaller particles that fit the voids created between neighboring larger particles~\cite{sohn68,brouwers2006,clusel2009}. For a detailed analysis of the RCPS formed by polydispersive particles, we divide these particles by size into six different spectra.  The first spectrum ($\Delta_{1}$) is a collection of particles with radii ranging from $1.0 \, \mu m$ to $2.0 \, \mu m$, the second spectrum ($\Delta_{2}$) is a collection of particles with radii ranging from from $2.0 \, \mu m$ to $3.0 \, \mu m$ and so on.  Figs.~\ref{fig:10a} and \ref{fig:10b} show the mean coordination number $z$ of polydispersive particles with random size distribution against the packing density for soft and hard particles, respectively. The insets give the mean coordination number $z$ in all spectra that compound the RCPS in the absence of long-range forces ($\varepsilon_{min}=0$). We realize, once again, the influence of the long-range forces over the packing process. At the end of the simulation, the mean coordination number for soft particles was $5.049 \pm 0.017$ for $\varepsilon_{min}=0.0$; $5.136 \pm 0.012$ for $\varepsilon_{min}=-5.0$ and $5.213 \pm 0.013$ for $\varepsilon_{min}=-10.0$. While for hard particles, we get $4.518 \pm 0.014$ for $\varepsilon_{min}=0.0$; $4.545 \pm 0.012$ for $\varepsilon_{min}=-5.0$ and $4.623 \pm 0.012$ for $\varepsilon_{min}=-10.0$. One should be aware that the precision of the different quantities calculated here might be overestimated, despite the few realizations considered, since the box imposes a spatial restriction over the initial non-overlapping particle size distribution, which leads to a biased ensemble. This effect is evidenced when one observes the spectral frequencies displayed in the insets of the RDF curves. The smaller particle is, the easier it is accepted inside the box.

Figs.~\ref{fig:11a} and \ref{fig:11b} display the time derivative of the kinetic energy of polydispersive particles as a function of time for soft and hard particles, respectively. The relaxation time is about $8.0 \, ms$ for soft particles and $6.0 \, ms$ for hard particles. It is seen that the time derivative of the kinetic energy is almost insensitive to long-range dispersive forces. Figs. \ref{fig:12}-\ref{fig:14} illustrate the  RDFs as defined by Eq.~\ref{eq:13} for each spectrum (spectral RDF) as a function of the radial distance considering different $\varepsilon_{min}$ values. For polydispersive particles, we set $\delta r_{i}=1.0 \, \mu \,m$ and $N_{r}=21$. The spectral RDFs for both soft (a) and hard particles (b) are shown in these figures. Insets in Figs.~\ref{fig:12}-\ref{fig:14} give the frequency of the particles for each spectrum. From these figures, one can see that the general shape of the spectral RDFs is practically unchanged by the long-range interaction forces, even though they strongly influence the transient state of the packing formation process. However, it is significantly changed when one increases the hardness of the particles. For soft particles, the first peaks of the spectral RDFs are sharper than those ones for hard particles. In other words, there is a higher probability of finding smaller particles layer around the bigger particles when these particles are softer. This behavior is more evident when one compares the corresponding RDFs of higher spectra for soft particles with those for hard particles.

\section{\label{sec:c} Conclusions}

MD simulations were performed to study the 2D random packing process of both monosized and random size fine particles within a micrometer scale. Both contact forces and long-range dispersive forces were taken into account in the simulations. Different physical quantities, including the packing density, mean coordination number, kinetic energy and RDF were calculated to study the packing process dynamics and characterize the formed RCPS over different values of both the long-range interaction strength and the hardness of the particles. It is found that the long-range forces can strongly influence the packing process dynamics as they might form large particle clusters, depending on the value of the long-range interaction strength. However, the general shape of the RDFs for the RCPS is seen to be more influenced by the hardness of the particles than by the long-range dispersive forces. Moreover, we obtained a high RCPS density for both monosized and polydispersive fine particles. As expected, the RCPS density for polydispersive particles are found to be slightly higher than that for monosized particles. Finally, we expect that this study can be helpful in the understanding of the packing process of fine particles with random size distribution as well as motivate experimental research on this matter.

\section{Acknowledgements}
We wish to thank UFERSA for computational support.




\bibliographystyle{model1a-num-names}

\end{document}